\documentclass[%
 reprint,
 amsmath,amssymb,
 aps,
prb
]{revtex4-1}
\usepackage{graphicx}
\usepackage{dcolumn}
\usepackage{bm}
\usepackage[dvipdfm,colorlinks,linkcolor=blue, urlcolor=blue, anchorcolor=blue, citecolor=blue]{hyperref}
\begin{document}
\title{Extended Dicke quantum battery with interatomic interactions and driving field}
\author{Fu-Quan Dou}
\email[]{doufq@nwnu.edu.cn}
\author{You-Qi Lu}%
\affiliation{
College of Physics and Electronic Engineering, Northwest Normal University,
Lanzhou, 730070, China}
\author{Yuan-Jin Wang}%
\affiliation{
College of Physics and Electronic Engineering, Northwest Normal University,
Lanzhou, 730070, China}%
\author{Jian-An Sun}%
\affiliation{
College of Physics and Electronic Engineering, Northwest Normal University,
Lanzhou, 730070, China}%

\begin{abstract}
We investigate the charging process of quantum battery (QB) systems in an extended Dicke model with both atomic interactions and an external driving field. We focus on the effects of the atomic interaction and the external driving field on the charging performance of QB 
and find that the maximum stored energy of QB has a critical phenomenon. We analyze the critical behavior and obtain the analytical expression of the critical atomic interaction. The dependence of the maximum stored energy, the energy quantum fluctuations and the maximum charging power on the number $N$ of the two-level systems are also discussed. In particular, for the maximum charging power, we obtain the quantum advantage of the QB, which approximately satisfies a superlinear scaling relation $P_{max}\propto N^{\alpha}$, where scaling exponent $\alpha$ varies with the number $N$ of the two-level systems. In the ultra-strong coupling regime, the atomic interaction can lead to a faster battery charging, and the quantum advantage $\alpha = 1.88$ can be achieved. While in the deep-strong coupling regime, the quantum advantage of the QB's maximum charging power is the same as that of the Dicke QB, i.e., $\alpha=1.5$.
\end{abstract}

\maketitle
\section{INTRODUCTION}
Miniaturization of electronic devices is inevitable with the development of industrial technology \cite{Millen_2016}. In the research field called ``quantum thermodynamics", the thermodynamics of small quantum systems has been considered in theoretical and experimental works \cite{Pekola2015,PhysRevA.99.062111}. In this context, one of the major issues triggered by potential technological applications is the possibility of efficiently storing energy in small systems exploiting quantum features and using it to provide power supply on-demand \cite{Carrega_2020}. It inspires the birth of quantum batteries (QBs), i.e., a quantum system that stores or supplies energy \cite{PhysRevE.87.042123,Campaioli2018,bhattacharjee2020quantum,Niedenzu2018,Giorgi2015}. The QB is based on quantum thermodynamics fundamentally different from traditional electrochemical batteries \cite{PhysRevApplied.14.024092,Skrzypczyk2014}.

Previous work have shown the importance of quantum resources in improving the performances of QBs,
such as work extraction \cite{PhysRevLett.111.240401,PhysRevLett.122.047702,e23050612,zhang2018enhanced}, charging power \cite{PhysRevLett.120.117702,PhysRevB.102.245407,Binder2015,PhysRevLett.118.150601}, and energy fluctuation \cite{Friis2018,PhysRevLett.125.040601,PhysRevE.98.032132,Perarnau_Llobet_2019,Crescente_2020,PhysRevResearch.2.023095}. In 2013 Alicki and Fannes first proposed that entangling unitary controls acting globally extract in general more work than unitary operations acting on each quantum cell separately \cite{PhysRevE.87.042123}. Entanglement generation can accelerate the process of work extraction, thereby leading to larger delivered power, which was first demonstrated in Ref. \cite{PhysRevLett.111.240401}. Afterward, two types of charging schemes were proposed \cite{Binder2015,PhysRevLett.118.150601}: parallel and collective charging scheme. In the collective charging scheme and for $N\geq2$, the charging power of a QB is larger than that in the parallel scheme \cite{Binder2015,PhysRevLett.118.150601,PhysRevLett.125.236402}. The collective scheme can achieve a speed-up in the charging process of QB, named ``quantum advantage" \cite{PhysRevLett.122.047702,zhang2018enhanced,PhysRevLett.120.117702,PhysRevB.102.245407,PhysRevLett.125.236402,Chen2020,PhysRevE.99.052106,yang2020,PhysRevE.103.042118,PhysRevA.97.022106,sen2019,PhysRevA.103.052220,PhysRevResearch.2.023113,PhysRevA.100.043833,PhysRevB.99.205437}. Besides, the performance of QB in the stable charging process \cite{PhysRevE.100.032107,PhysRevLett.124.130601,Dou2021,PhysRevE.101.062114,Mitchison2021chargingquantum}, self-discharging process \cite{PhysRevE.103.042118,Kamin_2020}, dissipation charging process \cite{PhysRevResearch.2.033413,PhysRevA.102.052223,PhysRevA.102.060201,PhysRevLett.122.210601,PhysRevResearch.2.013095}, and many-body interaction systems \cite{PhysRevA.97.022106,PhysRevB.100.115142,PhysRevA.101.032115} have been investigated.

In the quest for such quantum advantage and potential experimental implementations of QBs, QB has been proposed in various models, such as two-level systems (TLSs) \cite{Crescente_2020,Chen2020,PhysRevE.99.052106,yang2020,PhysRevE.103.042118}, three-level systems \cite{PhysRevE.100.032107,Dou2021,Dou2020}, quantum cavity model \cite{e23050612,zhang2018enhanced,PhysRevLett.120.117702,PhysRevB.102.245407,PhysRevB.98.205423,zhao2021,PhysRevA.103.033715,PhysRevB.99.035421}, interacting spin chain model \cite{PhysRevA.97.022106,PhysRevB.100.115142,PhysRevA.101.032115,PhysRevB.100.115142,PhysRevA.101.032115}, and Sachdev-Ye-Kitaev model \cite{PhysRevLett.125.236402,Rosa2020}.

The Dicke model, in which a collection of TLSs interacts with a single-photon cavity model, has extensive applications in quantum simulation, quantum sensing, quantum communication, and quantum computing \cite{PhysRev.93.99}. 
Novel findings---circuit quantum electrodynamics \cite{PhysRevA.69.062320} and solid-state semiconductor \cite{Gunter2009} have allowed the advent of the ultra-strong coupling (USC) regime \cite{Niemczyk2010,Bayer2017,FriskKockum2019,RevModPhys.91.025005,Macha2014,PhysRevLett.117.210503} and the deep-strong coupling (DSC) regime \cite{Yoshihara2017,PhysRevLett.105.263603,PhysRevA.96.013849,PhysRevA.102.013701,PhysRevLett.120.183601,PhysRevA.98.023863,PhysRevA.96.063821,PhysRevA.95.053824} 
in the Dicke model. Recently, the Dicke QB has been proposed \cite{PhysRevLett.120.117702}, in which the collective charging scheme can achieve a $\sqrt{N}$ speed-up in the charging process compared to the parallel charging scheme. In a recent experiment \cite{quach2020}, the Dicke QB was first implemented using an organic semiconductor as an ensemble of TLSs coupled to a confined optical mode in a microcavity \cite{PhysRevLett.120.117702}. In addition, a two-photon Dicke QB has also been introduced \cite{PhysRevB.99.205437}, constructed by coupling $N$ TLSs with a two-photon cavity mode. The two-photon interaction can lead to faster charging and higher averaged charging power (proportional to $N$ instead of $\sqrt{N}$) compared to the conventional Dicke QB.

Recently, an extended Dicke model that includes atomic interactions and an external driving field has been established for a Bose-Einstein condensate inside an ultra-high finesse optical cavity \cite{PhysRevA.78.023634}. Some exotic quantum phenomena arising from these unnoticed atom-atom interactions have been predicted or observed in theory and experiments, i.e., the second-order phase transition from the super radiation phase to the ``Mott" phase \cite{PhysRevA.78.023634,Chen_2009}. To take another step forward, one may naturally wonder if this extended Dicke model can further improve the performance of QBs.

\begin{figure}
\centering
\includegraphics[width=0.45\textwidth]{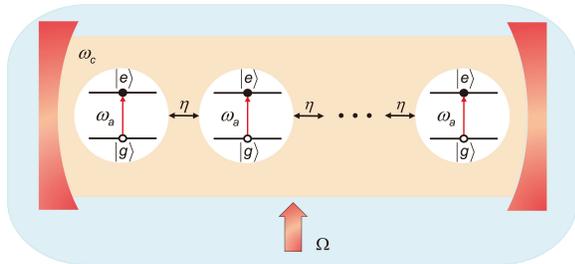}
\caption{Extended Dicke quantum battery. It includes a set of $N$ identical and two-level systems (TLSs) with a frequency $\omega_{a}$. TLSs have an atom-atom interaction with a strength of $\eta$. TLSs coupled with a single-photon cavity with a frequency of $\omega_{c}$ and an external driving field with a strength of $\Omega$. At $t=0$, each atom is in the ground state $|g\rangle$. At the period $T$, the quantum is fully charged, and the final state of atoms is $|e\rangle^{\otimes N}$.}
\label{fig1}
\end{figure}
In this paper, we introduce an extended Dicke model as a QB system, which includes atomic interactions and an external driving field. Here the battery consists of $N$ TLSs displayed in collective mode during the charging process, and the charger has the cavity, the atomic interaction and the external driving field. We compare with the two types of coupling in the charging process, i.e., the USC and the DSC regimes. We investigate the dependence of the stored energy, energy fluctuations, and the average charging power of the battery on the atomic interaction and external driving field. In addition, we introduce the quantum phase transition (QPT) to analyze the critical behavior of the maximum stored energy of the QB. We also analyze how the number $N$ of the TLSs influences the maximum stored energy, energy fluctuation and maximum charging power. Finally, we are primarily concerned with the quantum advantage of the maximum charging power of QB.

The rest of the paper is organized as follows. Section \ref{section2} introduces the extended Dicke QB, charging protocols and numerical approach. The influence of the atomic interactions and external driving fields on the stored energy, quantum fluctuations, and charging power is investigated in Section \ref{section3}. In Section \ref{section4}, we analyze the effects of the number $N$ of TLSs on the maximum storage energy, energy quantum fluctuation and maximum charging power of the quantum battery, and the quantum advantage of the maximum charging power of QB is discussed. Finally, we give a summary in Section \ref{section5}.
\section{MODEL AND APPROACH} \label{section2}
The QB model consists of a single-mode cavity field, $\emph{N}$ identical two-level atoms and an external driving field, as shown in Fig. \ref{fig1}. Here, we consider $N$ TLSs with infinite-range interactions coupled to a single cavity mode via a single-photon coupling \cite{PhysRevA.78.023634,PhysRevA.102.013701}. The total Hamiltonian of the QB system is (hereafter, we set $\hbar=1$)
\begin{equation}\label{Eq.1}
H(t)=H_{0}+\lambda(t)H_{1},
\end{equation}
where the time-dependent parameter $\lambda(t)$ describes the charging time interval, which we assume to be given by a step function equal to $1$ for $t\in[0,T]$ and zero elsewhere. $H_{0,1}$ are the Hamiltonians of the battery and charger, respectively, and with the following forms
\begin{equation}\label{Eq.2}
H_{0}=\omega_{a}\hat{J}_{z},
\end{equation}
\begin{eqnarray}\label{Eq.3}
\nonumber H_{1}&=&H_{c}+H_{a-c}+H_{a-a}+H_{field}\\
    &=&\omega_{c}\hat{a}^{\dag}\hat{a}+2\omega_{c}g\hat{J}_{x}(\hat{a}^{\dag}+\hat{a})+\frac{\eta}{N}\hat{J}_{z}^{2}+\Omega\hat{J}_{x},
\end{eqnarray}
Here, $\hat{a}(\hat{a}^{\dag})$ annihilates (creates) a cavity photon with frequency $\omega_{c}$ and $\hat{J}_\alpha=\frac{1}{2}\sum_{i}^{N}\sigma_{i}^{\alpha}$, with $\alpha=x,y,z$ as the components of a collective spin operator in terms of the Pauli operators $\sigma_{i}^{\alpha}$ of the $\emph{i}$-th TLS. The parameters $\omega_{a}$, $g$, $\eta$, and $\Omega$ denote the energy splitting between the ground $|g\rangle$ and excited state $|e\rangle$ of each TLS, the TLS-cavity coupling strength, the atomic interaction strength, and the driving field strength, respectively.

We consider the charging process of the extended Dicke QB in a closed quantum system. Here, the $N$ TLSs are prepared in ground state $|g\rangle$ and coupled to a single-mode cavity in the $N$ photon Fock state $|N\rangle$. Thus, the initial state of the total system is
\begin{equation}\label{Eq.4}
|\psi(0)\rangle=|N\rangle \otimes \underbrace{|g, \ldots, g\rangle}_{N} ,
\end{equation}
where $|\psi^{N}(0)\rangle$ is the initial state of the entire system. In our charging protocol, QB will start charging when the classical parameter $\lambda$ is nonzero. The wave function of the system evolves with time, i.e.,
\begin{equation}\label{Eq.5}
|\psi(t)\rangle=e^{-iHt}|\psi(0)\rangle.
\end{equation}

The stored energy of QB can be expressed in terms of the mean local energy of QB as follows:
\begin{equation}\label{Eq.6}
E(t)=\langle\psi(t)|H_{0}|\psi(t)\rangle-\langle\psi(0)|H_{0}|\psi(0)\rangle.
\end{equation}

Notably, the stored energy $E(t)$ that we are interested in does not include the Hamiltonian $H_{a-a}$ because the existence of atomic interactions promote positive or negative contributions in $E(t)$ and eventually lead to an uneven comparison \cite{PhysRevE.103.042118}. The average charging power of QB is defined by
\begin{equation}\label{Eq.7}
P(t)=\frac{E(t)}{t}.
\end{equation}

The stored energy and charging power are not sufficient to fully characterize QB. Then, we consider the energy quantum fluctuation as another useful quantifier of QB performance. To do so, we analyze the fluctuations between the initial and final time of the charging process represented by the correlator \cite{PhysRevB.102.245407,Friis2018,Crescente_2020}
\begin{eqnarray}\label{Eq.8}
\nonumber \Sigma^{2}(t)=&\Big[\sqrt{\langle H_{0}^{2}(t)\rangle-(\langle H_{0}(t)\rangle)^{2}}\\
                - &\sqrt{\langle H_{0}^{2}(0)\rangle-(\langle H_{0}(0)\rangle)^{2}}\Big]^{2}.
\end{eqnarray}

We emphasize that $\Sigma(t)$ is related to the inverse of the so-called reverse quantum speed limit, which can also be used to characterize the discharging of QB \cite{PhysRevA.104.042209}.
Because of the unitary evolution of the entire battery system during the charging process, the energy will be transferred back and forth between the charger and battery. Therefore, it is unnecessary to track the stored energy, charging power, and energy quantum fluctuation of QB at every moment. Usually, we choose the maximum stored energy $E_{max}$ (at time $t_{E}$), maximum charging power $P_{max}$ (at time $t_{P}$), and value of energy quantum fluctuation $\Sigma(t)$ at the time $t_{E}$, to measure QB performance,
\begin{eqnarray}\label{Eq.9}
E_{max}\equiv\max_{t}[E(t)]=E[(t_{E})],
\end{eqnarray}
\begin{eqnarray}\label{Eq.10}
P_{max}\equiv\max_{t}[P(t)]=P[(t_{P})],
\end{eqnarray}
\begin{eqnarray}\label{Eq.11}
\overline{\Sigma}\equiv\Sigma(t_{E}).
\end{eqnarray}

We focus on the resonance regime, i.e., $\omega_{a}=\omega_{c}$, to ensure the maximum energy transfer. We take $\omega_{a}$ as a dimensionless parameter and let $\omega_{a}=1$. Off-resonance case $\omega_{a} \neq \omega_{c}$ will not be discussed since they are characterized by a less efficient energy transfer between the cavity and TLSs \cite{Schleich2001,PhysRevB.102.245407}.

Similar to previous studies \cite{PhysRevLett.120.117702}, we must diagonalize Dicke Hamiltonian exactly to evaluate the stored energy, charging power, and energy quantum fluctuation. The reason is that the number of photons in the extended Dicke Hamiltonian is not conserved, which can be clearly seen from the interaction term $2\omega_{c}g\hat{J}_{x}(\hat{a}^{\dag}+\hat{a})$ of Eq. (\ref{Eq.3}) containing counter-rotating terms of the $\hat{a}^{\dagger}\hat{J}_+$, $\hat{a}\hat{J}_-$. However, we notice that $\hat{J}^{2}=\hat{J}_{x}^{2}+\hat{J}_{y}^{2}+\hat{J}_{z}^{2}$ is a conserved quantity for the Hamiltonian in Eq. (\ref{Eq.1}). A convenient basis set for representing the Hamiltonian is ${|n;j,m\rangle\equiv |n\rangle\otimes|j,m\rangle}$, where $\emph{n}$ indicates the number of photons, $j(j+1)$ is the eigenvalue of $\hat{J}^{2}$. Here, $\emph{m}$ denotes the eigenvalue of $\hat{J}_{z}$. Within this notation, the initial state in Eq. (\ref{Eq.2}) can be written as follows:
\begin{eqnarray}\label{Eq.12}
|\psi(0)\rangle=|N,N/2,-N/2\rangle.
\end{eqnarray}

The matrix elements of the Dicke Hamiltonian (\ref{Eq.1}) in the main text can be evaluated over the basis set $ |n,j,m\rangle$ using the following relations for ladder operator of photons and pseudo-spin
\cite{Miguel2011,PhysRevA.85.053831,PhysRevE.67.066203}
\begin{eqnarray}\label{Eq.13}
\hat{a}^{\dagger}|n,j,m\rangle=\sqrt{n+1}|n+1,j,m\rangle,
\end{eqnarray}
\begin{eqnarray}\label{Eq.14}
\hat{a}|n,j,m\rangle=\sqrt{n}|n-1,j,m\rangle,
\end{eqnarray}
\begin{eqnarray}\label{Eq.15}
\hat{J}_{\pm}|n, j, m\rangle=\sqrt{j(j+1)-m(m \pm 1)}|n, j, m \pm 1\rangle.
\end{eqnarray}
Notice that one can work in a subspace at fixed $j=N/2$ and read that $\hat{J}_{x}=(\hat{J}_{+}+\hat{J}_{-})/2$ because of the conservation of $\hat{J}^{2}$. Thus, we obtain
\begin{figure*}
\centering
\includegraphics[width=0.75\textwidth]{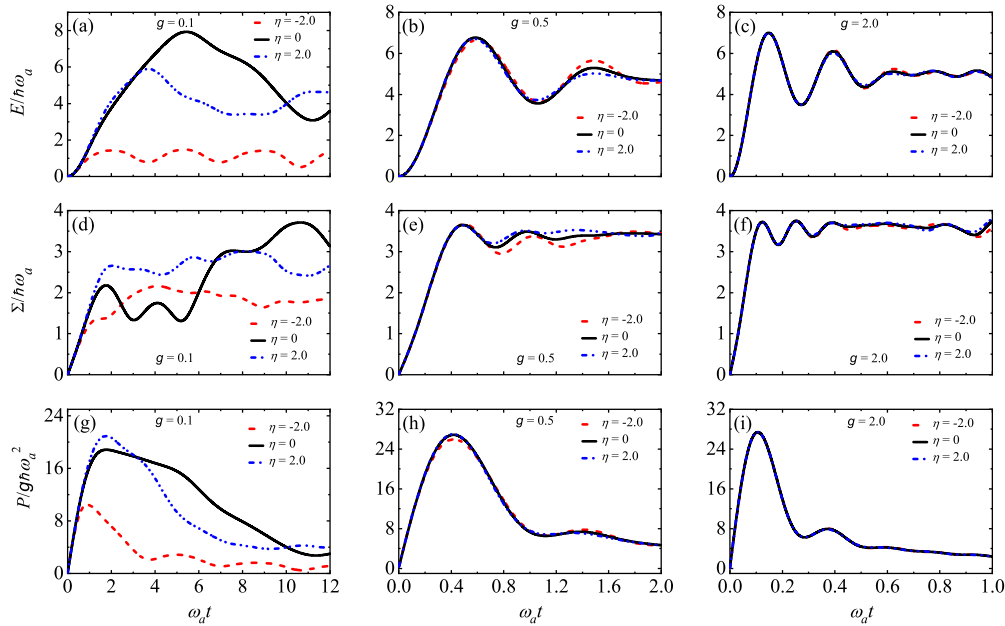}
\caption{(a)-(c) The dependence of the stored energy $E(t)$ (in units $\hbar \omega_{a}$), (d)-(f) energy quantum fluctuations $\Sigma(t)$ (in units $\hbar \omega_{a}$), and (g)-(i) average charging power $P(t)$ (in units $\hbar g \omega_{a}^{2}$) as a function of $\omega_{a}t$ for the different coupling regime $g$. The different curves in these plots stand for various $\eta$, as indicated in the legends, and all are for a QB with the number of TLSs $N=10$. We used different timescales in the panels to determine the maximum position of the various quantities in different cases.}
\label{fig2}
\end{figure*}
\begin{widetext}
\begin{eqnarray}\label{Eq.16}
\begin{aligned}
&\langle n^{\prime},\frac{N}{2},\frac{N}{2}-q^{\prime}|H|n,\frac{N}{2},\frac{N}{2}-q\rangle= \\
&\hbar\omega_{c}\{[n+\frac{N}{2}-q+\frac{\eta}{N}(\frac{N}{2}-q)^{2}]\delta_{n^{\prime},n} \delta_{q^{\prime},q}
+\Omega[f_{n,\frac{N}{2},\frac{N}{2}-q}^{(1)}\delta_{n^{\prime},n} \delta_{q^{\prime},q+1}+f_{n,\frac{N}{2},\frac{N}{5}-q}^{(2)}\delta_{n^{\prime},n} \delta_{q^{\prime},q-1}]\\
&+g[f_{n,\frac{N}{2},\frac{N}{2}-q}^{(3)}\delta_{n^{\prime},n+1} \delta_{q^{\prime},q+1}+f_{n,\frac{N}{2},\frac{N}{2}-q}^{(4)}\delta_{n^{\prime},n+1} \delta_{q^{\prime},q-1}
+f_{n,\frac{N}{2},\frac{N}{2}-q}^{(5)}\delta_{n^{\prime},n-1} \delta_{q^{\prime},q+1}+f_{n,\frac{N}{2},\frac{N}{2}-q}^{(6)}\delta_{n^{\prime},n-1} \delta_{q^{\prime},q-1}]\},
\end{aligned}
\end{eqnarray}
\end{widetext}
with
\begin{eqnarray}\label{Eq.17}
f_{k,j,m}^{(1)}=\sqrt{j(j+1)-m(m-1)},
\end{eqnarray}
\begin{eqnarray}\label{Eq.18}
f_{k,j,m}^{(2)}=\sqrt{j(j+1)-m(m+1)},
\end{eqnarray}
\begin{eqnarray}\label{Eq.19}
f_{k,j,m}^{(3)}=\sqrt{(k+1)[j(j+1)-m(m-1)]},
\end{eqnarray}
\begin{eqnarray}\label{Eq.20}
f_{k,j,m}^{(4)}=\sqrt{(k+1)[j(j+1)-m(m+1)]},
\end{eqnarray}
\begin{eqnarray}\label{Eq.21}
f_{k,j,m}^{(5)}=\sqrt{k[j(j+1)-m(m-1)]},
\end{eqnarray}
\begin{eqnarray}\label{Eq.22}
f_{k,j,m}^{(6)}=\sqrt{k[j(j+1)-m(m-1)]}.
\end{eqnarray}

We remark that the number of photons is not conserved by the Dicke Hamiltonian. It is also not bounded from above; thus, it may take an arbitrarily large integer value. In practice, we need to introduce a cutoff $N_{ph} > N$ on the maximum number $N_{ph}$ of photons within our finite-size numerical diagonalization. This choice allows us to select a case scenario of large $N$ values to calculate the stored energy without making any significant difference. In the following, we show numerical results obtained from an exact numerical diagonalization scheme for $N=1,...,30$. We have examined that excellent numerical convergence is achieved by selecting the maximum number of photons as $N_{ph}=4N$ \cite{PhysRevLett.120.117702,PhysRevB.102.245407}.
\section{THE CHARGING PROPERTY} \label{section3}
In this section, we discuss the charging property of the extended Dicke QB. We will present an analysis for different coupling strengths, ranging from the USC regime ($0.1 \leq g < 1.0$) to the DSC regime ($ g \geq 1.0$). We expect faster charging and further enhancement of the average charging power compared to that observed in the single-photon Dicke QB \cite{PhysRevLett.120.117702}. 
\subsection{Charging properties with interatomic interactions}
We first analyze the charging properties of QB only with atomic interactions. In this case, the overall system can be described by the following Hamiltonian
\begin{eqnarray}\label{Eq.23}
H=\omega_{c}\hat{a}^{\dag}\hat{a}+\omega_{a}\hat{J}_{z}+2\omega_{c}g\hat{J}_{x}(\hat{a}^{\dag}+\hat{a})+\frac{\eta}{N}\hat{J}_{z}^{2}.
\end{eqnarray}

Fig. \ref{fig2} illustrates the time evolution of the stored energy, energy quantum fluctuations and average charging power for various values of atomic interaction strengths in different coupling regimes. The blue dash-dotted lines and red dashed lines indicate the repulsive ($\eta>0$) and attractive ($\eta<0$) interactions, respectively. For comparison purposes, the charging process without atomic interaction is also depicted with black solid lines.

When the battery system is in a weak USC regime, the atomic interaction (whether repulsive or attractive) always has a negative impact on the energy storage of the QB. However, for the average charging power, different atomic interactions have different effects, in which the repulsive interaction increases and the attractive interaction decreases the maximum charging power. As the coupling strength increases, the influence of the atomic interaction becomes weaker. When the coupling strength increases to the DSC regime, the stored energy and the average charging power of the QB are almost unaffected. This represents an interesting fact of the intrinsic competition between the atom-atom and atom-cavity field interactions \cite{PhysRevA.78.023634}.

\renewcommand{\tabcolsep}{0.18cm}
\renewcommand{\arraystretch}{1.3}
\begin{table}[!htbp]
  \centering
  \caption{Maximum stored energy $E(t_{E})$ (in units of $N \omega_{a}$) and energy quantum fluctuation $\overline{\Sigma}$ (in units of $N \omega_{a}$) of the corresponding charging time $t_{E}$ for the USC and DSC regimes for $N=10$.}
      \begin{tabular}{lllllll}
      \hline\hline
             & \multicolumn{2}{c}{$g=0.1$} & \multicolumn{2}{c}{$g=0.5$} & \multicolumn{2}{c}{$g=2.0$} \\
             \cline{2-7}
             & \multicolumn{1}{c}{$E(t_{E})$} & \multicolumn{1}{c}{$\overline{\Sigma}$} & \multicolumn{1}{c}{$E(t_{E})$} & \multicolumn{1}{c}{$\overline{\Sigma}$} & \multicolumn{1}{c}{$E(t_{E})$} & \multicolumn{1}{c}{$\overline{\Sigma}$} \\ \hline
      \textit{$\eta=-2.0$} &1.473  &2.013  &6.662  &3.468  &6.992  &3.533  \\
      \textit{$\eta=0$}    &7.931  &1.391  &6.768  &3.473  &7.000  &3.528  \\
      \textit{$\eta=2.0$}  &5.899  &2.525  &6.709  &3.488  &6.995  &3.530 \\ \hline\hline
      \end{tabular}%
  \label{TAB.1}%
\end{table}
\begin{figure}
\centering
\includegraphics[width=0.45\textwidth]{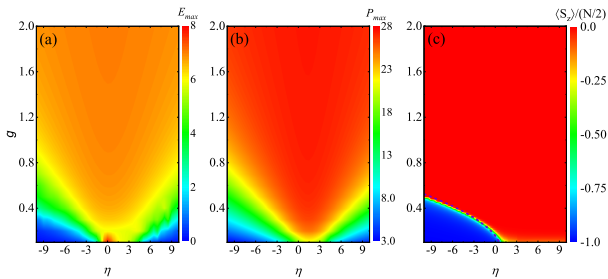}
\caption{(a) and (b) Contour plots of QB's maximum stored energy $E_{max}$ (in units $\hbar \omega_{a}$)  and charging power $P_{max}$ (in units $\hbar g \omega_{a}^{2}$) as functions of the coupling strength $g$ and interatomic interactions strength $\eta$. (c) Phase diagram described by $\langle S_{z}\rangle /(N/2)$ as functions of the atomic interaction strength $\eta$ and coupling strength $g$. The dashed line is the critical curve of the ``Mott'' phase (for $|\langle S_{z}\rangle /(N/2)|=0$) and normal phase (for $|\langle S_{z}\rangle /(N/2)|=1$). All plots correspond to the case of $N=10$.}
\label{fig3}
\end{figure}
The evolution of quantum fluctuations of energy with time and the value of energy quantum fluctuation $\Sigma(t)$ at the time $t_E$ when the maximum stored energy occurs is shown in Fig. \ref{fig2}(d)-(f) and Table \ref{TAB.1}, respectively. The energy fluctuations are high in all cases considered, expect for the case of the without interactions in the weak USC regime, where a better charge is achieved, and we therefore get $\overline{\Sigma}\sim 1.391$. As a result, the QB is not fully charged in the two coupling regimes, i.e., the DSC and USC regimes, due to the internal interaction between the $N$ TLSs.

We then calculate the maximum stored energy and the maximum charging power of the QB as a function of the coupling strength and the atomic interaction strength shown in Fig. \ref{fig3}. In the USC regime, the atomic interaction can significantly change the QB's maximal stored energy and charging power. However, in the DSC regime, this effect is almost negligible. In particular, the maximum stored energy of the QB has a critical behavior in the USC regime, i.e., the system exists a critical point and the maximum stored energy of the QB changes obviously near the critical point.

To further understand the critical phenomenon, we introduce the QPT \cite{Chen_2006,Baumann2010,Brennecke2007,Li2013}. In fact, this system has two independent QPT parameters, i.e., the atomic interaction strength $\eta$ and the coupling strength $g$. The extended Dicke model only with the atomic interaction has a second-order quantum phase transition from the super-radiant phase to the ``Mott'' phase \cite{PhysRevA.78.023634,Chen_2009}, and the critical value of the atomic interaction strength $\eta$ at the QPT point has the following expression (dashed line in Fig. \ref{fig3} (c))
\begin{eqnarray}\label{Eq.24}
\eta = \omega_{a}-\frac{4g^{2}N}{\omega_{c}}.
\end{eqnarray}
For $\eta=0$, the Hamiltonian (\ref{Eq.23}) reduces to the Dicke model. At the critical point $g=\sqrt{\omega_{a} \omega_{c}}/2$, the system undergoes a well-known normal-superradiant phase transition \cite{PhysRevE.67.066203}. The phase diagram of the scaled the proportional inversion of the TLS $\langle S_{z}\rangle /(N/2)$ with respect to the independent QPT parameters \cite{PhysRevE.67.066203} is illustrated in Fig. \ref{fig3} (c). The critical curve (dashed line in Fig. \ref{fig3} (c)) in the phase diagram appears as the intersection of the two phase regimes of the ``Mott'' (for $|\langle S_{z}\rangle /(N/2)|=0$) and normal (for $|\langle S_{z}\rangle /(N/2)|=1$) phases \cite{PhysRevE.99.052106}. Correspondingly, the maximum stored energy of the QB changes significantly at the critical point of the QPT.
\begin{figure*}
\centering
\includegraphics[width=0.75\textwidth]{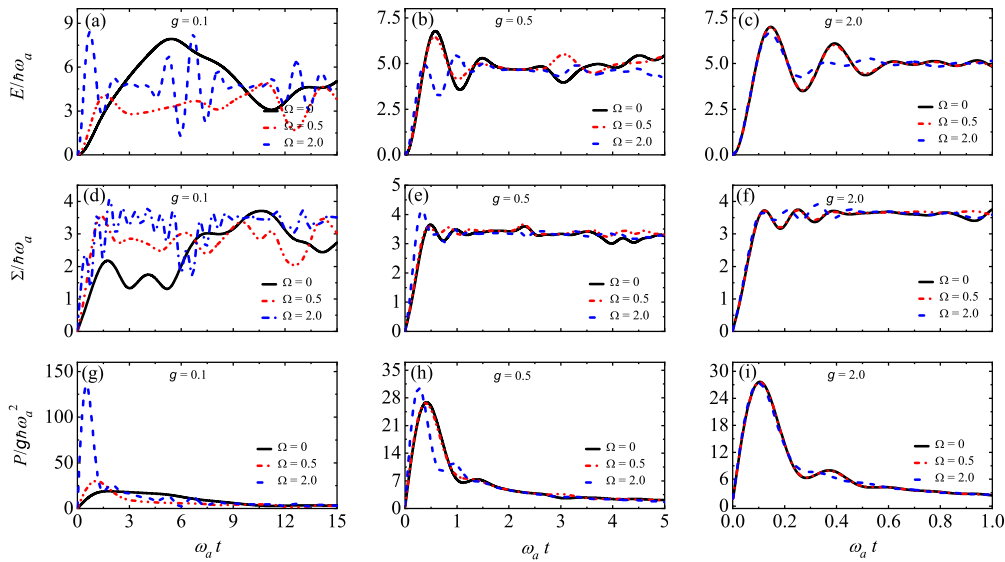}
\caption{(a)-(c) The dependence of the stored energy $E(t)$ (in units $\hbar \omega_{a}$), (d)-(f) energy quantum fluctuations $\Sigma(t)$ (in units $\hbar \omega_{a}$), and (g)-(i) average charging power $P(t)$ (in units $\hbar g \omega_{a}^{2}$) as a function of $\omega_{a}t$ with the number of TLSs $N=10$ for the different coupling regimes $g$. The different curves in these plots stand for various of $\Omega$, as indicated in the legends, and all are for a QB with the number of TLSs $N=10$. We used different timescales in the panels to determine the maximum position of the various quantities in different cases.}
\label{fig4}
\end{figure*}
\subsection{Charging properties with external driving field}
In this section, we investigate the charging properties of QB only with an external driving field. In this case, the overall system can be described by the extended Dicke Hamiltonian as follows:
\begin{eqnarray}\label{Eq.25}
H=\omega_{c}\hat{a}^{\dag}\hat{a}+\omega_{a}\hat{J}_{z}+2\omega_{c}g\hat{J}_{x}(\hat{a}^{\dag}+\hat{a})+\Omega\hat{J}_{x}.
\end{eqnarray}

In Fig. \ref{fig4}, we illustrate the time evolution of the stored energy, the energy quantum fluctuation, and the average charging power for various value of the driving field strength in different coupling regimes. In order to analyze the benefits of the external driving field, we also plot the situation without external driving field (solid black lines), corresponding to $\Omega=0$. In the weak USC regime, the driving field increases the stored energy of the QB. In particular, with the increase of the driving field strength, the QB requires less time to achieve the maximum stored energy. This also increases the maximum charging power of the QB (see Fig. \ref{fig4}(g)). However, in the DSC regime, the advantages of the external driving field become smaller and smaller.

As in Figs. \ref{fig4} (d)-(f) and Table \ref{TAB.2}, energy quantum fluctuations are unavoidable and finite in all considered parameter ranges. This result is directly related to the fact that the QB is not fully charged, due to the interaction between the $N$ TLSs and the cavity mode.
\renewcommand{\tabcolsep}{0.20cm}
\renewcommand{\arraystretch}{1.3}
\begin{table}[!htbp]
  \centering
  \caption{Maximum stored energy $E(t_{E})$ (in units of $N \omega_{a}$) and its quantum fluctuation $\overline{\Sigma}$ (in units of $N \omega_{a}$) of the corresponding charging time $t_{E}$ for the USC and DSC regimes for $N=10$.}
      \begin{tabular}{lllllll}
      \hline\hline
             & \multicolumn{2}{c}{$g=0.1$} & \multicolumn{2}{c}{$g=0.5$} & \multicolumn{2}{c}{$g=2.0$} \\
             \cline{2-7}
             & \multicolumn{1}{c}{$E(t_{E})$} & \multicolumn{1}{c}{$\overline{\Sigma}$} & \multicolumn{1}{c}{$E(t_{E})$} & \multicolumn{1}{c}{$\overline{\Sigma}$} & \multicolumn{1}{c}{$E(t_{E})$} & \multicolumn{1}{c}{$\overline{\Sigma}$} \\ \hline
      \textit{$\Omega=0$}        &7.931  &1.391   &6.768  &3.473   &7.000  &3.528  \\
      \textit{$\Omega=0.5$}      &4.917  &2.944   &6.451  &3.421   &6.978  &3.522  \\
      \textit{$\Omega=2.0$}      &8.424  &1.443   &5.437  &3.443   &6.667  &3.486 \\ \hline\hline
      \end{tabular}%
  \label{TAB.2}%
\end{table}

\begin{figure}[!htbp]
\centering
\includegraphics[width=0.45\textwidth]{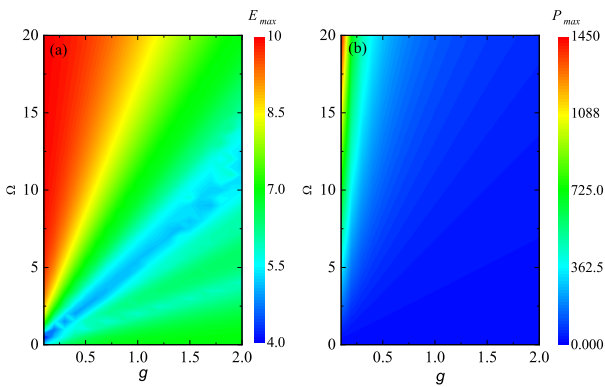}
\caption{Contour plots of QB's maximum stored energy and charging power as functions of the coupling regimes $g$ and the external driving strength $\Omega$. (a) QB's maximum stored energy $E_{max}$ (in units $\hbar \omega_{a}$) and (b) maximum charging power $P_{max}$ (in units $\hbar g \omega_{a}^{2}$). All plots correspond to the case of $N=10$.}
\label{fig5}
\end{figure}

To further demonstrate the impact of the external driving field on the performance of the QB in different coupling regimes, we calculate the maximum stored energy and charging power as a function of the coupling strength and the external driving strength for a fixed number of TLSs $N$ (see in Fig. \ref{fig5}). The external driving field increases the maximum stored energy and charging power of the QB. Particularly, in the USC regime, with the increase of the driving field strength, the QB has a higher maximum charging power and a greater maximum stored energy compared to the DSC regime. More interestingly, the maximum stored energy of the QB always oscillates in a fixed region. When the battery system is in the USC regime, an external driving field brings QB closer to the maximum stored energy. Therefore, in view of actual experimental implementations, it is necessary to properly design an extended Dicke QB driven by an external field in the USC regime to achieve QB with larger stored energy and higher average charging power.
\begin{figure*}[!ht]
\centering
\includegraphics[width=0.8\textwidth]{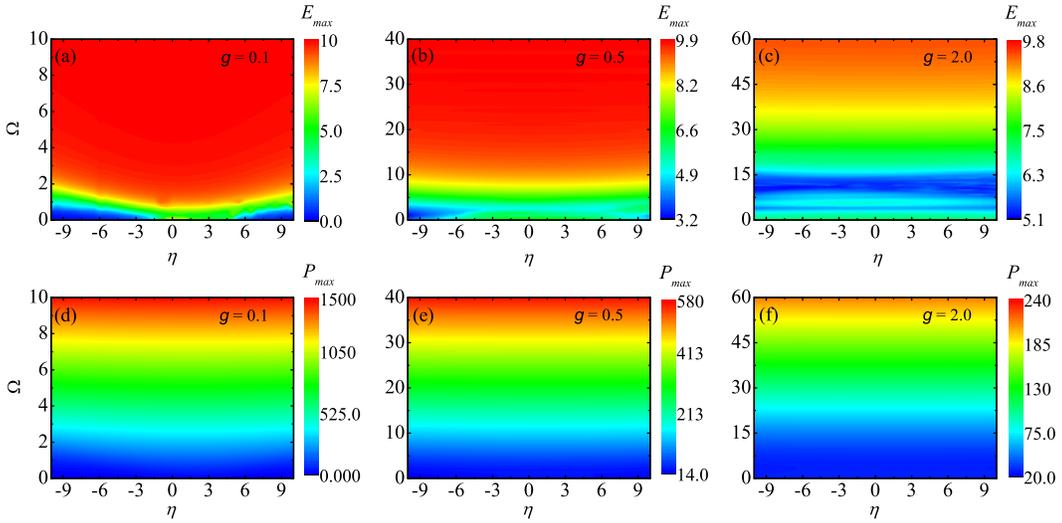}
\caption{(a)-(c) Contour plots of QB's maximum stored energy $E_{max}$ (in units $\hbar \omega_{a}$), (d)-(f) contour plots of QB's maximum charging power $P_{max}$ (in units $\hbar g\omega_{a}^{2}$) as functions of the atomic interaction strength $\eta$ and driving field strength $\Omega$ for different coupling regimes $g$. All plots correspond to the case of $N=10$.}
\label{fig6}
\end{figure*}
\subsection{Charging properties with atomic interactions and external driving field}
In order to further discuss the combined effects of both the atomic interaction and the external driving field on the charging process, in this section, we calculate the maximum stored energy and charging power as a function of both the atomic interactions and external driving field for the various coupling regimes as shown in Fig. \ref{fig6}.

\begin{figure*}[!ht]
\centering
\includegraphics[width=0.85\textwidth]{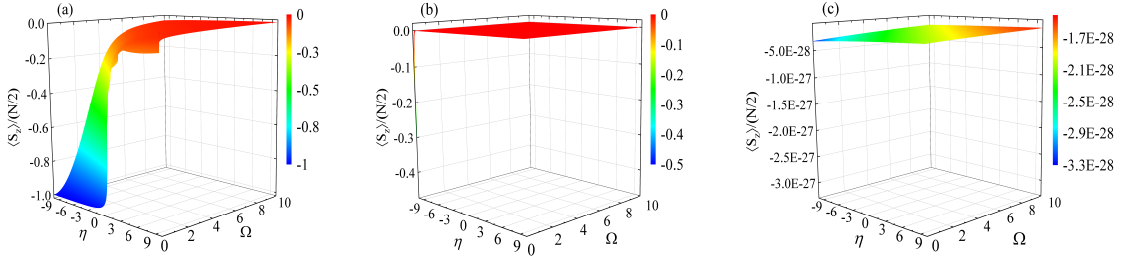}
\caption{Phase diagrams described by the scaled population inversion of the TLSs $\langle S_{z}\rangle/ (N/2)$ as a function of the atomic interaction strength $\eta$ and the driving field strength $\Omega$ for different coupling regimes (a) $g=0.1$, (b) $g=0.5$ and (c) $g=2.0$. Other parameters are the same as these in Fig. \ref{fig6}.}
\label{fig7}
\end{figure*}
In the weak USC regime, if the driving field strength is weak, the atomic interaction has a great influence on the maximum stored energy of the QB. However, with the increase of the driving field strength, the maximum stored energy of the QB is largely unaffected by the atomic interaction (see Figs. \ref{fig6} (a)-(c)). This reflects the competition between the external driving field and the atomic interaction. In particularly, compared with the DSC regime, the driving field strength required for the QB to obtain the maximum stored energy in the USC regime is smaller. For the maximum charging power, no matter what the coupling regime is, the effect of the atomic interaction on the maximum charging power of QB can be neglected, and the external drive field increases the maximum charging power. In particular, the QB in the USC regime also has a higher maximum charging power compared to the DSC regime (see Fig. \ref{fig6} (d)-(f)).

It is more interesting that, in this case, the maximum stored energy of the QB also has a critical behavior with the weak driving field strength. Similarly, we calculate the QPT. The phase diagrams of the scale population inversion of the TLSs $\langle S_{z}\rangle/ (N/2)$ as functions of the atomic interaction strength and the driving field strength in different coupling regimes are illustrated in Fig. \ref{fig7}. In the USC regime, the QPT induced by the atomic interaction always occurs, while it does not occur within the parameter range we considered in the DSC regime. As a result, when the system is at the QPT point, the atomic interaction will greatly influence the maximum stored energy.
\section{ADVANTAGE OF COLLECTIVE CHARGING}\label{section4}
For a classical battery device, the electric current is static so that a charging process can be completed in a certain time. However, in the case of the quantum battery, the energy transfer is subject to dynamic development and depends on the devices and the charging time \cite{PhysRevA.103.052220}. We have investigated how the charging process depends on the charging time and internal parameters when the number of charger and battery are fixed. Let us now find out how the charging process depends on the number of batteries (TLSs) when the charger is fixed. We have set the initial state of the QB $|\psi(0)\rangle = |n;j,m\rangle$, which leads to the dynamical evolution of the battery involving highly entangled Dicke states and a collective charging of the Dicke QB.

\begin{figure*}
\centering
\includegraphics[width=0.85\textwidth]{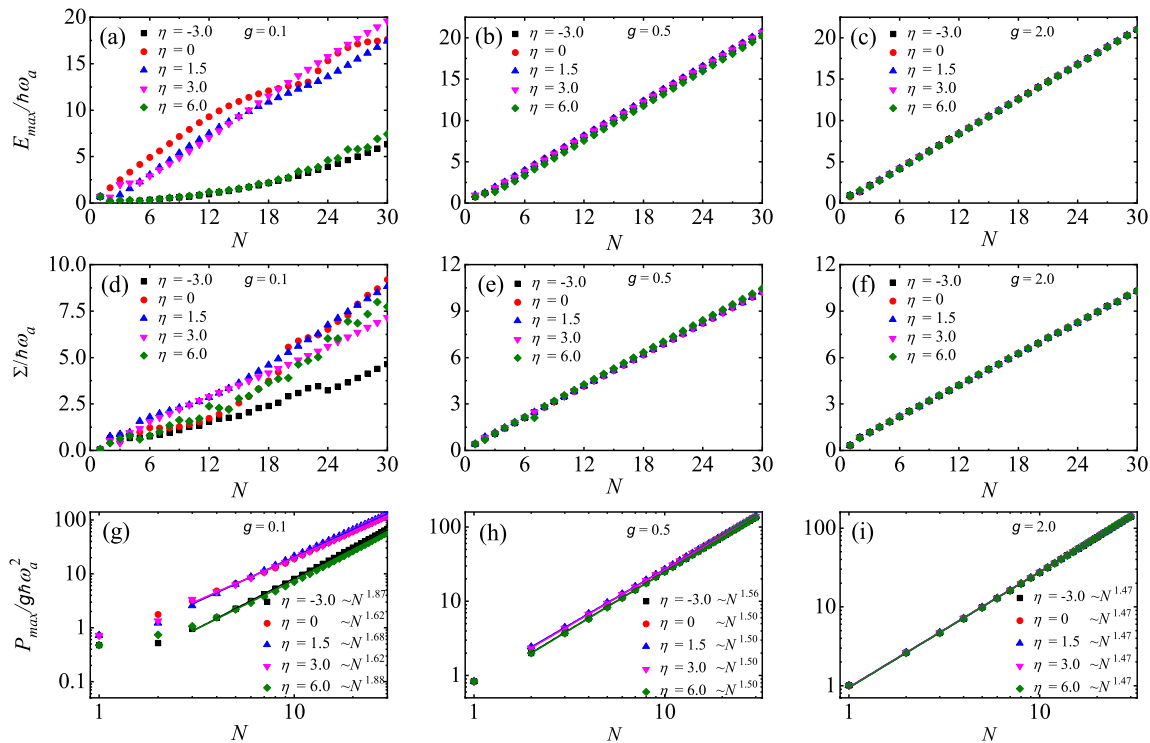}
\caption{(a)-(c) Maximum stored energy $E_{max}$ (in units $\hbar \omega_{a}$), (d)-(f) the value of the
energy fluctuations at the maximum energy $\overline{\Sigma}$ (in units $\hbar \omega_{a}$) and (g)-(i) maximum charging power $P_{max}$ (in units $\hbar g\omega_{a}^{2}$) of the QB with the number $N$ of TLSs for different coupling regimes $g$ and atomic interactions $\eta$. The solid lines in panels (g)-(i) indicate the numerical fitting of the power scaling (\ref{Eq.26}) in a logarithmic scale for $N\in[1,30]$.}
\label{fig8}
\end{figure*}
We analyze the maximum stored energy $E_{max}$ in Eq. (\ref{Eq.5}), the maximum charging power $P_{max}$ in Eq. (\ref{Eq.6}), and the value of the energy quantum fluctuations $\overline{\Sigma}$ at maximum stored energy in Eq. (\ref{Eq.8}) as a function of the number $N$ of the TLSs. We recall that in Ref. \cite{PhysRevLett.120.117702}, it was shown that the energy scales with $N$, whereas the average charging power scales like $P\propto N^{3/2}$ in single-photon Dicke QB for large $N$. Therefore, we naturally expect the existence of a general scaling relation between the charging power of the extended Dicke QB and the number $N$ of TLSs. We assume that the maximum charging power takes the following form
\begin{equation}\label{Eq.26}
P_{max}\propto \beta N^{\alpha}.
\end{equation}

By taking the logarithm, we use linear fitting to obtain the scaling exponent $\alpha$
\begin{equation}\label{Eq.27}
\log(P_{max}) = \alpha \log(N) + \log(\beta).
\end{equation}
The scaling exponent $\alpha$ essentially reflects the collective nature of the battery in transferring energy.
\subsection{Collective charging with atomic interactions}
In Fig. \ref{fig8}, we display the QB's maximum stored energy, energy quantum fluctuations and maximum charging power as a function of the number $N$ of the TLSs for different atomic interactions ($\eta=-3.0$ (black squares), $\eta=1.5$ (blue up-triangles), $\eta=3.0$ (magenta down-triangles) and $\eta=6.0$ (olive diamonds)) in different coupling regimes. For comparison purposes, the case without the atomic interaction is also depicted with red circles ($\eta=0$).

Similarly, in the weak USC regime the atomic interaction has a great influence on the maximum stored energy and the energy fluctuation of the QB (see in Figs. \ref{fig8} (a)-(f)). Compared to the case with  the atomic interaction, the QB always has a higher maximum stored energy in a small number $N$ of TLSs without interatomic interaction. As the coupling strength increases, the influence of the atomic interaction becomes weaker. When the coupling strength increases to the DSC regime, the effect of the atomic interaction can almost be neglected, and the maximum stored energy and energy fluctuation always increase linearly with the number $N$ of TLSs.

\renewcommand{\tabcolsep}{0.3cm}
\renewcommand{\arraystretch}{1}
\begin{table}[!htbp]
  \centering
  \caption{Battery power scaling exponent $\alpha$ obtained from numerical fitting of Eq. (\ref{Eq.26}) vs. the number of TLSs $N$ for different atomic interactions $\eta$ and coupling strengths $g$.}
      \begin{tabular}{lllllll}
      \hline\hline
             & \multicolumn{2}{c}{$g=0.1$} & \multicolumn{2}{c}{$g=0.5$} & \multicolumn{2}{c}{$g=2.0$} \\
             \cline{2-7}
            & \multicolumn{1}{c}{$\alpha_{1}$} & \multicolumn{1}{c}{$\beta_{1}$}
            & \multicolumn{1}{c}{$\alpha_{2}$} & \multicolumn{1}{c}{$\beta_{2}$}
            & \multicolumn{1}{c}{$\alpha_{3}$} & \multicolumn{1}{c}{$\beta_{3}$} \\ \hline
      \textit{$\eta=-3.0$} &1.87  &0.38  &1.56  &0.85  &1.47  &0.97  \\
      \textit{$\eta=0$}    &1.62  &0.73  &1.50  &0.93  &1.46  &0.98  \\
      \textit{$\eta=1.5$}  &1.68  &0.70  &1.50  &0.92  &1.46  &0.98  \\
      \textit{$\eta=3.0$}  &1.62  &0.73  &1.50  &0.93  &1.46  &0.98  \\
      \textit{$\eta=6.0$}  &1.88  &0.36  &1.56  &0.84  &1.47  &0.97  \\ \hline\hline
      \end{tabular}%
  \label{TAB.3}%
\end{table}%
\begin{figure}[!htbp]
\centering
\includegraphics[width=0.4\textwidth]{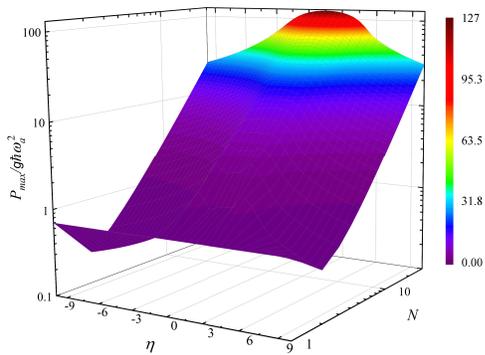}
\caption{Logarithmic contour plot of the maximum charging power with the number $N$ of TLSs and atomic interaction $\eta$. It shows different values of power scaling exponent $\alpha$ in different atomic interactions. We set $g=0.1$, and consider the ranges $N\in[1,30]$ and $\eta\in[-10,10]$.}
\label{fig9}
\end{figure}

We are interested in the general scaling relationship between the maximum charging power and the number $N$ of TLSs. As shown in Figs. \ref{fig8} (g)-(i), the logarithmic plot of the maximum charging power $P_{max}$ gives the scaling exponent $\alpha$ for the region $N\in[1,30]$, see Table \ref{TAB.3}. In the weak USC regime, the scaling exponent can be close to $\alpha \rightarrow {1.88}$. While as the coupling strength increases, the maximum charging power scaling exponent of the battery decreases. When the coupling strength increases into the DSC regime, this scaling exponent decreases to $\alpha \rightarrow {1.5}$, which is consistent with the result of Dicke QB \cite{PhysRevLett.120.117702}. Interestingly, with $\eta =1.5$ as the centre, the maximum charging power scaling exponent will increase symmetrically as $\eta$ deviates from $1.5$ in both directions (see table \ref{TAB.3}).

\begin{figure}
\centering
\includegraphics[width=0.45\textwidth]{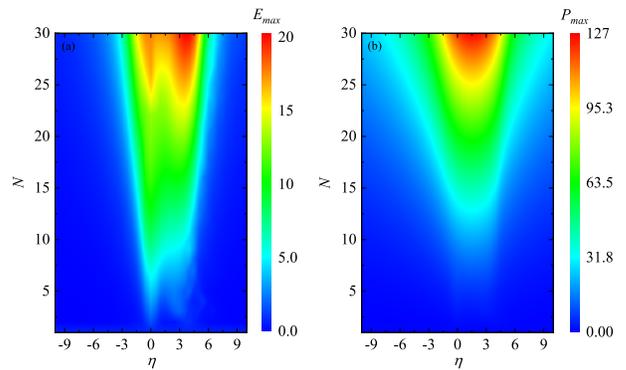}
\caption{Contour plot of the maximum stored energy and charging power as a function of the atomic interactions $\eta$ and the number $N$ of TLSs. (a) QB's maximum stored energy $E_{max}$ (in units $\hbar\omega_{a}$) and (b) the maximum charging power $P_{max}$ (in units $\hbar g\omega_{a}^{2}$). We set $g=0.1$ and consider the ranges $N\in[1,30]$ and $\eta\in[-10,10]$.}
\label{fig10}
\end{figure}

It is worth noting that in the weak USC regime, the atomic interaction has a great influence on the maximum charging power scaling exponent $\alpha$. Fig. \ref{fig9} shows the scaling exponent $\alpha$ of the maximum charging power with respect to the numbers of TLSs $N$ for different atomic interactions. Here, we set the coupling strength $g=0.1$ and consider the ranges $N\in[1,30]$ and $\eta\in[-10,10]$. This figure confirms the observations in Fig. \ref{fig8}(g). For a fixed value of the atomic interaction $\eta$, the slope of the upward plane indicates the value of $\alpha$. In order to further prove the effects of the atomic interaction and the number $N$ of TLSs on the maximum stored energy and the maximum charging power of the QB in the weak USC regime. In Fig. \ref{fig10} we calculate the maximum stored energy and charging power of the QB as a function of the atomic interaction and the number $N$ of TLSs for fixed coupling strength. The atomic interaction plays a negative role in the QB's stored energy for a small number $N$ of TLSs. However, for a large but finite value of the number of $N$, the place where the QB's maximum stored energy occurs shifts from near no atomic interaction the weak repulsive interaction range, i.e., the ranges $\eta\in[2.5,4.5]$ observed from the figure (see in Fig. \ref{fig10}(a)). While for the maximum charging power, a weak repulsive interaction increases the maximum charging power of the QB, and the maximum charging power is roughly symmetrical with respect to around $\eta = 1.5$.
\begin{figure*}
\centering
\includegraphics[width=0.8\textwidth]{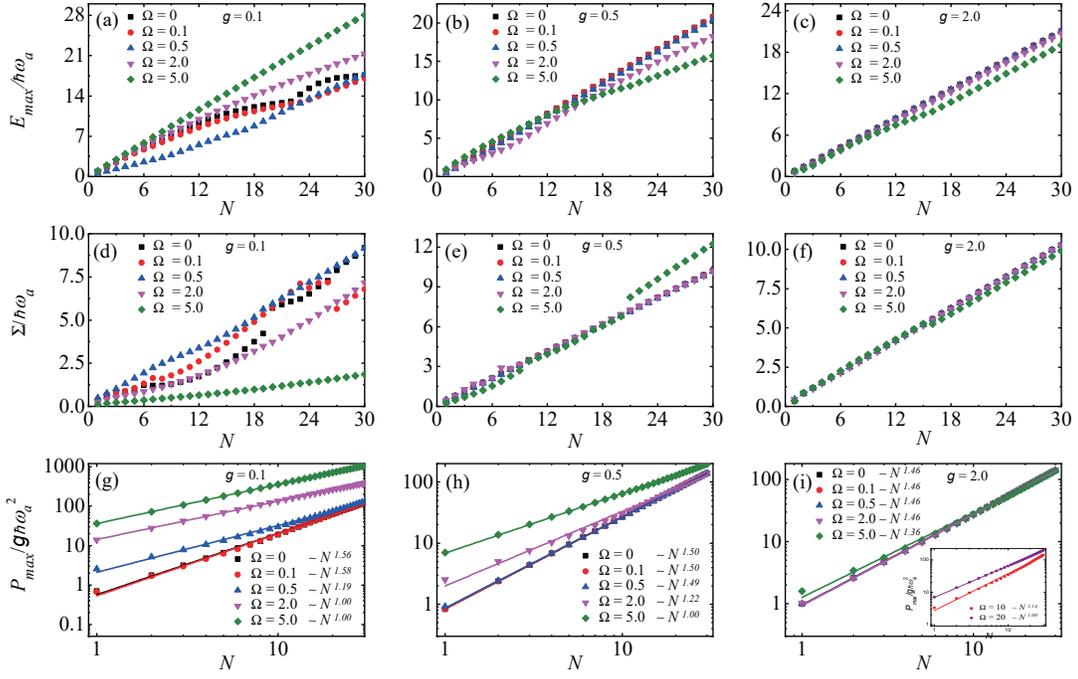}
\caption{(a)-(c) Maximum stored energy $E_{max}$ (in unit $\hbar \omega_{a}$), (d)-(f) the value of the
energy fluctuations at the maximum energy $\overline{\Sigma}$ (in unit $\hbar \omega_{a}$) and (g)-(i) the maximum charging power $P_{max}$ (in unit $\hbar g \omega_{a}^{2}$) of the QB with the number $N$ of TLSs for different coupling regimes $g$ and external driving strengths $\Omega$. The solid lines in panel (g)-(i) represent the numerical fitting of the power scaling (\ref{Eq.26}) in a logarithmic scale for $N\in[1,30]$.}
\label{fig11}
\end{figure*}
\subsection{Collective charging with external driving field}
For the case of only external driving field, we also calculate the maximum stored energy, the energy quantum fluctuations and the maximum charging power as a function of the number $N$ of TLSs for different external driving field strengths in different coupling regimes in Fig. \ref{fig11}. Here, the driving field strengths is $\Omega=0$ (black squares), $\Omega=0.1$ (red circles), $\Omega=0.5$ (blue up-triangles), $\Omega=2.0$ (magenta down-triangles) and $\Omega=5.0$ (olive diamonds). No matter what the coupling strength is, the maximum stored energy and energy fluctuation of the QB always increase linearly for the number $N$ of TLSs (see Fig. \ref{fig11} (a)-(f)). In particular in the USC regime, the external driving field can make QB reach full charging.

\renewcommand{\tabcolsep}{0.3cm}
\renewcommand{\arraystretch}{1.2}
\begin{table}[!htbp]
  \centering
  \caption{Battery power scaling exponent $\alpha$ obtained from numerical fitting of Eq. (\ref{Eq.26}) with the number $N$ of TLSs for different coupling strengths $g$ and external driving strengths $\Omega$.}
      \begin{tabular}{lllllll}
      \hline\hline
             & \multicolumn{2}{c}{$g=0.1$} & \multicolumn{2}{c}{$g=0.5$} & \multicolumn{2}{c}{$g=2.0$} \\
             \cline{2-7}
            & \multicolumn{1}{c}{$\alpha_{1}$} & \multicolumn{1}{c}{$\beta_{1}$}
            & \multicolumn{1}{c}{$\alpha_{2}$} & \multicolumn{1}{c}{$\beta_{2}$}
            & \multicolumn{1}{c}{$\alpha_{3}$} & \multicolumn{1}{c}{$\beta_{3}$} \\ \hline
      \textit{$\Omega=0$}   &1.56  &0.78  &1.50  &0.93  &1.46  &0.98  \\
      \textit{$\Omega=0.1$} &1.58  &0.76  &1.50  &0.93  &1.46  &0.98  \\
      \textit{$\Omega=0.5$} &1.19  &1.38  &1.49  &0.94  &1.46  &0.98  \\
      \textit{$\Omega=2.0$} &1.00  &3.17  &1.22  &1.27  &1.46  &0.98  \\
      \textit{$\Omega=5.0$} &1.00  &4.76  &1.00  &2.15  &1.36  &1.10  \\ \hline\hline
      \end{tabular}%
  \label{TAB.4}%
\end{table}%
The logarithmic plot of the maximum power $P_{max}$ gives the scaling exponent $\alpha$ for the number $N$ of TLSs, as shown in Fig. \ref{fig11} (g)-(i) and Table \ref{TAB.4}. Regardless of whether the USC regime or DSC regime, the external driving field increases the maximum charging power of the QB. However, with the increase of the external driving field, the scaling exponent of the maximum charging power finally gradually $\alpha \rightarrow 1$. We find that this scaling exponent becomes $\alpha \rightarrow 1$ when the driving field strength is $10$ times the coupling strength, i.e., $\Omega \geq10 g$.

\begin{figure}
\centering
\includegraphics[width=0.4\textwidth]{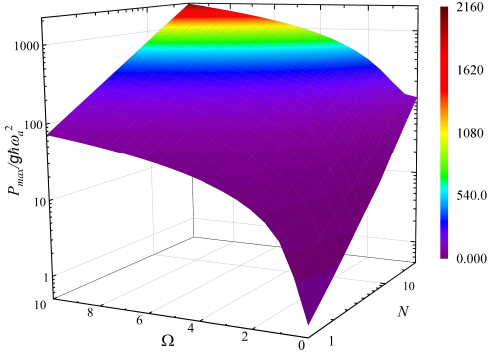}
\caption{Logarithmic contour plot of the maximum charging power vs. the number $N$ of TLSs and external driving strengths $\Omega$. It shows different values of power scaling exponent $\alpha$ in different external driving strengths. We set $g=0.1$, and consider the ranges $N\in[1,30]$ and $\Omega\in[0,10]$.}
\label{fig12}
\end{figure}
\begin{figure}
\centering
\includegraphics[width=0.45\textwidth]{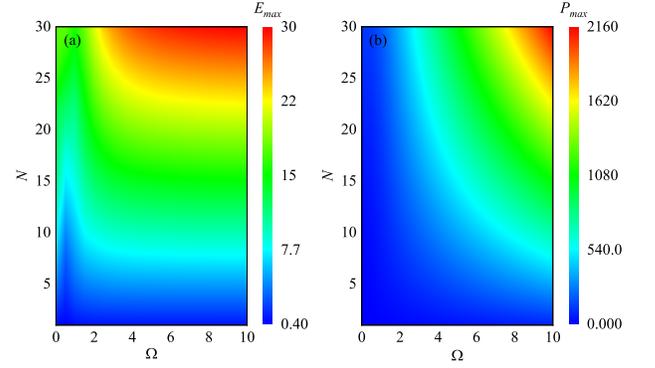}
\caption{Contour plot of the maximum stored energy and charging power as a function of the external driving strengths $\Omega$ and the number $N$ of TLSs. (a) The QB's maximum stored energy $E_{max}$ (in unit $\hbar\omega_{a}$) and (b) the maximum charging power $P_{max}$ (in unit $\hbar g\omega_{a}^{2}$). Here, we set $g=0.1$ and consider the ranges $N\in[1,30]$ and $\Omega\in[1,10]$.}
\label{fig13}
\end{figure}

Similarly, we further demonstrate the scaling exponent of the maximum charging power with respect to the number $N$ of TLSs for different external driving fields in the USC regime (see in Fig. \ref{fig12}). 
For a fixed value of the external driving field, the slope of the upward plane indicates the value of $\alpha$. This scaling exponent $\alpha$ stabilizes with the increase of the external driving field strength gradually, and this is consistent with the results obtained in Fig. \ref{fig11} (g) and Table \ref{TAB.4}. Moreover, we also illustrate the dependence of the maximum stored energy and maximum charging power on the external driving strength and the number of TLSs $N$. The external driving field and the number $N$ of the TLSs increase the maximum stored energy and the maximum charging power (see in Fig. \ref{fig13}).

\begin{figure*}
\centering
\includegraphics[width=0.8\textwidth]{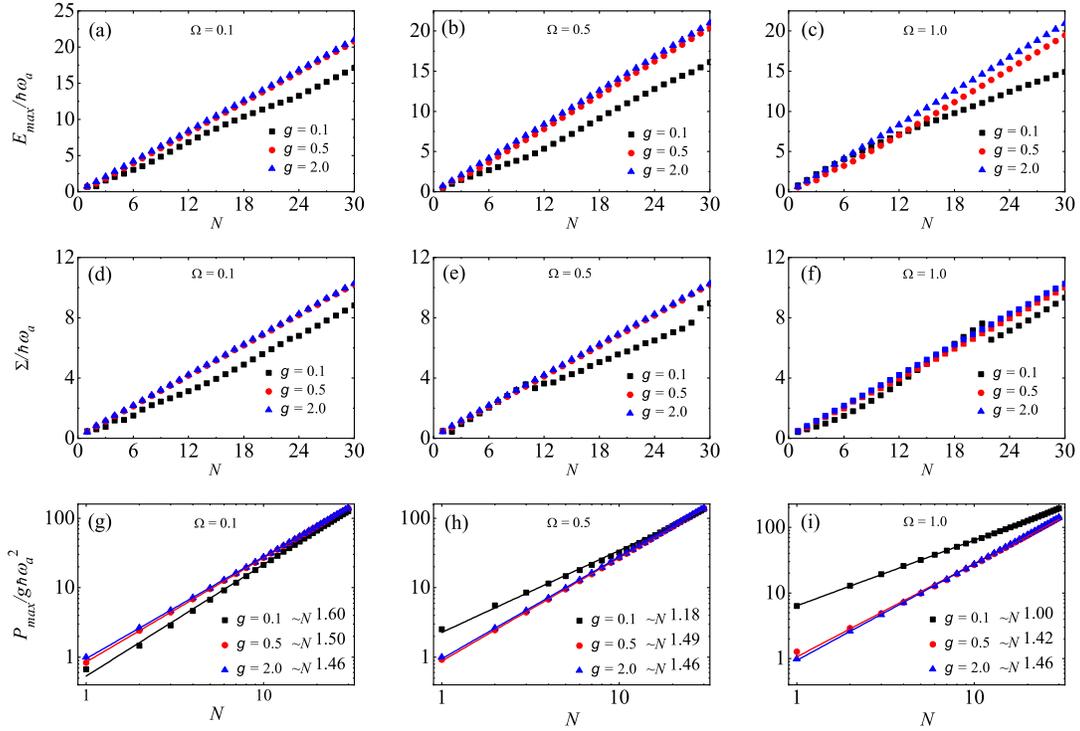}
\caption{(a)-(c) Maximum stored energy $E_{max}$ (in unit $\hbar \omega_{a}$), (d)-(f) the value of the
energy fluctuations at the maximum of the energy $\overline{\Sigma}$ (in unit $\hbar \omega_{a}$) and (g)-(i) the maximum charging power $P_{max}$ (in unit $\hbar g \omega_{a}^{2}$) of the QB with the number $N$ of TLSs for different coupling regimes $g$ and external driving strengths $\Omega$. The solid lines in panel (d)-(f) show the numerical fitting of the power relation (\ref{Eq.26}) in logarithmic scale for $N\in[1,30]$. Another parameter is $\eta=1.5$.}
\label{fig14}
\end{figure*}
\subsection{Collective charging with both atomic interactions and external driving field}
This section further discusses how the charging process depends on the number of TLSs in the case with both the atomic interactions and external driving fields. Fig. \ref{fig14} illustrates the calculation of the maximum energy, the energy quantum fluctuations and the maximum charging power as a function of the number $N$ of TLSs in different coupling regimes.

As the previous results show, if there is only atomic interaction, the weak repulsive interaction leads to the maximum charging power of the quantum battery. While if there exists only an external drive field, the external drive field increases the maximum charging power of the quantum battery, and the scaling exponent of the maximum charging power decreases to $\alpha=1$. In order to obtain the QB with a larger stored energy, a higher charging power and a better scaling exponent, we therefore select the atomic interaction strengths and driving field strengths appropriately, i.e., we set the atomic interaction strength $\eta=1.5$ and the external driving strength $\Omega=0.1$, $\Omega=0.5$, and $\Omega=1.0$.
\renewcommand{\tabcolsep}{0.3cm}
\renewcommand{\arraystretch}{1.35}
\begin{table}[!htbp]
  \centering
  \caption{Scaling exponent $\alpha$ obtained from the numerical fitting of Eq. (\ref{Eq.26}) with the number $N$ of TLSs for various values $g$ and $\Omega$.}
      \begin{tabular}{lllllll}
      \hline\hline
             & \multicolumn{2}{c}{$g=0.1$} & \multicolumn{2}{c}{$g=0.5$} & \multicolumn{2}{c}{$g=2.0$} \\
             \cline{2-7}
            & \multicolumn{1}{c}{$\alpha_{1}$} & \multicolumn{1}{c}{$\beta_{1}$}
            & \multicolumn{1}{c}{$\alpha_{2}$} & \multicolumn{1}{c}{$\beta_{2}$}
            & \multicolumn{1}{c}{$\alpha_{3}$} & \multicolumn{1}{c}{$\beta_{3}$} \\ \hline
      \textit{$\Omega=0.1$} &1.60  &0.38  &1.18  &0.85  &1.00  &0.97  \\
      \textit{$\Omega=0.5$} &1.50  &0.46  &1.49  &0.90  &1.42  &0.97  \\
      \textit{$\Omega=1.0$} &1.46  &0.73  &1.46  &0.93  &1.46  &0.98  \\ \hline\hline
      \end{tabular}%
  \label{TAB.5}%
\end{table}%

No matter what the coupling regime is, the maximum stored energy and the maximum charging power of the QB increase linearly with the number $ N $ of TLS (see Fig. \ref{fig14} (a)-(f)). For the maximum charging power, we are more interested in the scaling exponent of the maximum charging power. We find that the effects of the atomic interaction and the coupling strength on the scaling exponent of the maximum charging power are different. In particular with weak driving field strength and coupling strength, i.e., $g=0.1$, and $\Omega=0.1$, this scaling relationship can reach $\alpha = 1.60$ (see Fig. \ref{fig14} (g)-(i))), which is higher than that of Dicke QB \cite{PhysRevLett.120.117702}.
\section{Conclusions} \label{section5}
We have introduced the concept of extended Dicke quantum battery, consisting of an array of entangled two-level systems with both atomic interactions and external driving fields. We have analyzed the influence of atomic interactions and external driving fields on the performance of QB in different coupling regimes, including the stored energy, energy quantum fluctuations, and the average charging power. We have demonstrated that in the weak USC regime, the atomic interaction (whether repulsive or attractive) always has a negative effect on the stored energy of the QB. The repulsive interaction increases the maximum charging power, and the attractive interaction decreases it. However, no matter what the coupling regime is, the external driving field increases the maximum stored energy and charging power of the QB. Moreover, we have found that the maximum stored energy always exists a critical behavior. The critical atomic interaction and coupling strengths have been obtained analytically. We have also investigated the effect of the number $N$ of TLSs on the QB's maximum stored energy, energy quantum fluctuations and maximum charging power in different coupling regimes. No matter whether the atomic interaction or the external driving field, the QB's maximum stored energy and energy quantum fluctuation increase linearly with the $N$ TLSs in different coupling regimes. In particular, we have obtained the quantum advantage of the maximum charging power of the QB, which approximately satisfies a scaling relation $P_{max}\propto N^{\alpha}$. In the USC regime, the atomic interaction, for a finite-size system, can lead to a higher average charging power (scaling exponent of the maximum charging power $\alpha=1.88$) compared to the Dicke QB. While in the DSC regime, the quantum advantage of the QBs maximum charging power is the same as that of the Dicke QB ($\alpha=1.5$). For the case of the external driving field, no matter what the coupling regime is, the external driving field strength increases the maximum charging power of the QB. Nevertheless, with the increase of the driving field strength, the advantage of the maximum charging power becomes weak (scaling exponent of the maximum charging power can only reach $\alpha=1$).

Recently, experimental efforts have been devoted to quantum simulations of an array of two-level systems, with, for example, a solid-state platform \cite{PhysRevLett.120.117702,Forn2017}, trapped ions \cite{PhysRevX.8.021027}, and cold atoms \cite{Baumann2010}, which could be considered as quantum battery. Our QB charging model can also be realized physically. For instance, we can choose an experimental setup for a BEC of $^{87}$Rb atoms coupled to a QED cavity \cite{Brennecke2007,PhysRevA.78.023634}. The BEC with two levels $|1\rangle$ and $|2\rangle$ is prepared in a time-averaged, orbiting potential magnetic trap. After moving the BEC into an ultrahigh-finesse optical cavity, an external controllable classical laser is applied to produce various transitions of the atoms between $|1\rangle$ and $|2\rangle$ states, and thus the charging of the QB could be implemented realistically.

\begin{acknowledgments}
The work is supported by the National Natural Science Foundation of China (Grant No. 12075193).
\end{acknowledgments}



\bibliography{reference}
\end{document}